\documentclass[twocolumn,showpacs,superscriptaddress,amsmath,amssymb]{revtex4}
\usepackage{graphicx}
\usepackage{dcolumn}
\usepackage{bm}
\usepackage{color}
\usepackage{mathrsfs}

\def\bd{
\begin{document}} \def\ed{\end{document}}
\def\bmp{\begin{minipage}} \def\emp{\end{minipage}}
\def\bcc{\begin{center}} \def\ecc{\end{center}}     \def\npg{\newpage}
\def\beq{\begin{equation}} \def\eeq{\end{equation}} \def\hph{\hphantom}
\def\be{\begin{equation}} \def\ee{\end{equation}} \def\r#1{$^{[#1]}$}
\def\n{\noindent} \def\ni{\noindent} \def\pa{\parindent}
\def\hs{\hskip} \def\vs{\vskip} \def\hf{\hfill} \def\ej{\vfill\eject}
\def\cl{\centerline} \def\ob{\obeylines}  \def\ls{\leftskip}
\def\underbar#1{$\setbox0=\hbox{#1} \dp0=1.5pt \mathsurround=0pt
   \underline{\box0}$}   \def\ub{\underbar}    \def\ul{\underline}
\def\f{\left} \def\g{\right} \def\e{{\rm e}} \def\o{\over} \def\d{{\rm d}}
\def\vf{\varphi} \def\pl{\partial} \def\cov{{\rm cov}} \def\ch{{\rm ch}}
\def\la{\langle} \def\ra{\rangle} \def\EE{e$^+$e$^-$} \def\pt{p_{\rm t}}
\def\pti{p_{{\rm t},i}} \def\vti{v_{{\rm t},i}}
\def\ptj{p_{{\rm t},j}}\def\Pt{P_{\rm t}} \def\vt{v_{\rm t}}

\def\bitz{\begin{itemize}} \def\eitz{\end{itemize}}
\def\btbl{\begin{tabular}} \def\etbl{\end{tabular}}
\def\btbb{\begin{tabbing}} \def\etbb{\end{tabbing}}
\def\beqar{\begin{eqnarray}} \def\eeqar{\end{eqnarray}}
\def\\{\hfill\break} \def\dit{\item{-}} \def\i{\item}
\def\bbb{\begin{thebibliography}{9} \itemsep=-1mm}
\def\ebb{\end{thebibliography}} \def\bb{\bibitem}
\def\bpic{\begin{picture}(260,240)} \def\epic{\end{picture}}
\def\akgt{\cl{\bf ACKNOWLEDGMENTS}}
\def\fgn{\noindent{\bf\large\bf figure captions}}
\def\m1{\langle N_p\rangle} \def\u2{\langle N_{\bar p}\rangle} \def\Nap{N_{\bar
p}}
\def\lan{\langle}
\def\ran{\rangle}
\def\p{\pi}
\def\ifmath#1{\relax\ifmmode #1\else $#1$\fi}%
\def\rc{\ifmath{{\mathrm{c}}}}
\def\cut{\ifmath{{\mathrm{cut}}}}
\def\rF{\ifmath{{\mathrm{F}}}}
\def\rK{\ifmath{{\mathrm{K}}}}
\def\rp{\ifmath{{\mathrm{p}}}}
\def\rt{\ifmath{{\mathrm{t}}}}
\def\LAB{\ifmath{{\mathrm{LAB}}}}
\def\cut{\ifmath{{\mathrm{cut}}}}
\def\beq{\begin{equation}}
\def\eeq{\end{equation}}

\newcommand{\cinst}[2]{$^{\mathrm{#1}}$~#2\par}
\newcommand{\crefi}[1]{$^{\mathrm{#1}}$}
\newcommand{\crefii}[2]{$^{\mathrm{#1,#2}}$}
\newcommand{\crefiii}[3]{$^{\mathrm{#1,#2,#3}}$}
\newcommand{\HRule}{\rule{0.5\linewidth}{0.5mm}}

\bd
\title{ Influence of initial size on higher cumulant ratios of net-proton number fluctuations}

\author{Fengbo Xiong} 
\affiliation{Key Laboratory of Quark and Lepton Physics (MOE) and
Institute of Particle Physics, Central China Normal University, Wuhan 430079, China}
\author{Lizhu Chen} 
\affiliation{Key Laboratory of Quark and Lepton Physics (MOE) and
Institute of Particle Physics, Central China Normal University, Wuhan 430079, China}
\author{Lin Li} 
\affiliation{Key Laboratory of Quark and Lepton Physics (MOE) and
Institute of Particle Physics, Central China Normal University, Wuhan 430079, China}
\author{Zhiming Li}
\affiliation{Key Laboratory of Quark and Lepton Physics (MOE) and
Institute of Particle Physics, Central China Normal University, Wuhan 430079, China}
\author{Yuanfang Wu} 
\affiliation{Key Laboratory of Quark and Lepton Physics (MOE) and
Institute of Particle Physics, Central China Normal University, Wuhan 430079, China}

\begin{abstract}
With the help of AMPT default model, we study the influence of initial size (centrality of collisions) on higher cumulant ratios of net-proton distributions. If the centrality is presented by impact parameter, there is a strong centrality dependent, in particular, in those peripheral collisions. This dependence is slightly reduced if the centrality is presented by number of participant, or charged multiplicity. However, the dynamical ratios are almost centrality independent. So the centrality dependence of dynamical ratios at RHIC beam energy scan are presented.

\end{abstract}


\maketitle
\section{Introduction}

One of the main goals of relativistic heavy ion collisions is to map the QCD phase diagram~\cite{Star-BES},
in particular, the critical point. Recently, the higher cumulants ratios of net-proton distribution
are suggested to be a good probe of QCD critical point
(QCP) ~\cite{Stephanov-prl91, Stephanov-prl102, Asakawa-prl103, Star-prl, Stephanov-prl107}.
They are skewness and kurtosis defined as,
\begin{equation}
S= \frac{\langle\delta N^3\rangle}{\langle\delta N^2\rangle^{3/2}},
\hspace{0.8cm}
\kappa= \frac{\langle\delta N^4\rangle}{\langle\delta N^2\rangle^2}-3.
\end{equation}
Where $\delta N = N - \langle N \rangle $, and $N$ is the number of net-proton. The skewness and the kurtosis present respectively
the asymmetry and the peak of the distributions. For a Gaussian distribution, they are zero. While near the QCD critical point,
a non-Gaussian distribution is expected. They deviate from zero, and change the sign~\cite{Asakawa-prl103, Stephanov-prl107}.

However, it is found afterwords that those ratios are influenced by various trivial effects,
such as the centrality bin width~\cite{weight}, and statistical fluctuations due to the finite number of proton
produced at RHIC incident energies~\cite{bialas, Lizhu-jpg}.
To eliminate influence of centrality bin width, the centrality weighted method are recommended~\cite{weight}.
To subtract the statistical fluctuations,  the dynamic ratios are suggested,
\begin{equation}
S_{\rm dyn}= S - S_{\rm stat},
\hspace{0.8cm}
\kappa_{\rm dyn}= \kappa - \kappa_{stat}.
\end{equation}
Where statistical parts of proton and anti-proton are determined by Poisson-like fluctuations.
The statistical fluctuations of net-proton is given by a Skellam-like distribution.
So, $S_{\rm stat}= \frac{\langle N\rangle}{\langle M\rangle^{3/2}}$ and $\kappa_{\rm stat}= \frac{1}{\langle M\rangle}$.
Here, $M$ is the mean of  total proton number.They are both zeros if proton and anti-proton are emitted as independent Poisson.

In addition to the above mentioned effects, the initial size, i.e., the centrality of collisions, should also influence the measurements.
The centrality of collisions is originally defined by impact parameter ($b$). It is the distance of two incident nuclei.
It takes minimum and maximum in central and most peripheral collisions, respectively. However, it is not experimentally measurable.
In experiment, the number of participating nucleons ($N_{\rm part}$), and charged multiplicity ($N_{\rm ch}$)~\cite{centrality} are usually used
to quantify the centrality.
In fact, only charged multiplicity of final state is directly measurable. From any one of these three variables,
the other two variables can be obtained by the Monte Carlo simulation of Glaube model. Conventionally, the centrality in model
calculations are classified according to the impact parameter. While in real data analysis,
it is usually given by the number of participants, which is deduced from directly measured multiplicity.

The influence of centrality definitions varies from observable to observable. Some observables are not sensitive to the definitions,
like elliptic flow $v_2$. But others are not,  such as forward-backward correlations, and system size. Their centrality dependences change significantly if the definition of centrality changes from one to another~\cite{Konchakovski}. In this paper, we will study the influence of centrality definitions on higher cumulant ratios using the AMPT default model~\cite{ampt}.

\section{Influence of centrality definitions }

In AMPT default model~\cite{ampt}, we can easily classify the centrality by impact parameter, number of participants, and multiplicity, respectively.
They are usually quantified by percentile. So in Fig.~1, we present the centrality dependence of Kurtosis (a), dynamical kurtosis (b), $s\sigma$(c), and $\kappa\sigma^2$ (d) for Au + Au collisions at 39GeV, where centralities are defined by impact parameter (red open stars), number of participants (blue open crosses), and multiplicity (black open triangles) in $|\eta|< 0.5$, respectively. All ratios are calculated in the same phase space cuts as RHIC/STAR data~\cite{Star-prl}. The centrality weighted method is also used here to eliminate the effects of centrality bin width.

\begin{figure*}
\includegraphics[width=0.9\textwidth]{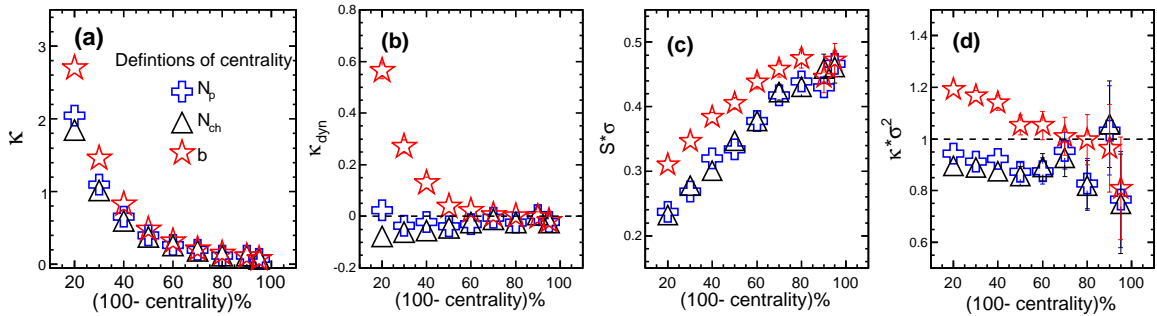}
\caption{Centrality dependence of kurtosis (a), dynamic kurtosis (b), S*$\sigma$ (c) and $\kappa$*$\sigma^2$ (d) for Au + Au collisions at 39 GeV, where centrality are classified by $N_{\rm part}$(blue open crosses), $N_{\rm ch}$(black open triangles), and $b$(red open stars), respectively.\label{fig1}}
\end{figure*}

We can see from 4 sub-figures that red open stars are obviously higher than both blue open crosses and black open triangles,
in particular, in peripheral collisions. While later two points are close to each other. This shows if the size fluctuations are presented by impact parameter, it contributes more to the ratios. The ratios are sensitive to the initial size fluctuations. So whenever we compare the data with model calculations, the same centrality definition should be used.

It is interesting to notice that centrality dependence of dynamical kurtosis in Fig.~1(b) are well reduced if the centrality is defined by number of participants, or multiplicity. Moreover, they are close to zero in all centralities, in contrary to the kurtosis given in Fig~1(a). So dynamical ratios well reduce the influence of the initial size fluctuations.

\section{Energy and centrality dependences of dynamical ratios }

As we know, there is no critical fluctuations in AMPT default model.  The results from the model can serve as a good reference of background effects. So the energy and centrality dependences of dynamical skewness and kurtosis for Au + Au collisions at seven RHIC/BES incident energies, 7.7, 11.5, 19.6, 27, 39, 62.4, 200GeV, are presented in Fig.~2(a) and (b), respectively. Where the centrality is defined by $N_{\rm ch}$, and plotted by $N_{\rm part}$, same as experimental data.

\begin{figure}
\includegraphics[width=0.48\textwidth]{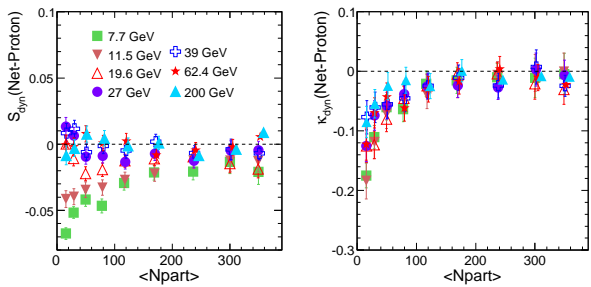}
\caption{Centrality dependence of dynamical skewness (a) and kurtosis (b) for Au + Au collisions at 7 RHIC/BES energies}
\end{figure}

We can see from Fig.~2(a) that at 2 lowest incident energies, the dynamical skewness are negative in peripheral collisions and becomes smaller and smaller towards the peripheral collisions. At 5 higher incident energies, they are around zero and almost energy and centrality independent. The dynamical kurtosis in Fig.~2(b) are incident energy independent. They are all negative in peripheral collisions, become smaller and smaller towards peripheral collisions, and are close to zero in mid-central to central collisions. So in peripheral collisions, there are still contributions in additional to Poisson-like statistical fluctuations in the measurements. The other effects, such as cuts of phase space and so on, may
contribute to it too, and related studies are on going.

\section{Summary}

With the help of the AMPT default model, we study how the initial size (centrality of collisions) influence the higher cumulant ratios of net-proton. It is found that if the centrality are presented by number of participant and multiplicity, their corresponding centrality dependence are similar. But they are quite different from that which centrality is presented by impact parameter. So when we compare the experimental data with model calculations, the same centrality definition should be used.
It is also found that dynamical ratios well reduce the influence of initial size. So the centrality dependence of dynamical skewness and kurtosis
for Au + Au collisions at seven RHIC/BES energies are presented. They can served as a good reference of background effects.

\section{Acknowledgments}
This work is supported in part by the NSFC of China with project No. 10835005 and 11221504.

\ed